\documentclass[pra,aps,showpacs,twocolumn,floatfix]{revtex4-1}
\usepackage[utf8]{inputenc}
\usepackage{graphicx}
\usepackage{amsmath}
\usepackage{amsfonts}
\usepackage{braket}
\usepackage{hyperref}
\usepackage{txfonts}
\newcommand{\expected}[1]{\left\langle#1\right\rangle}

\let\Re\relax

\DeclareMathOperator{\Re}{Re}

\newcommand{\Tr}{\text{Tr}}
\begin{document}
\title{Enhanced estimation of loss in the presence of Kerr nonlinearity}
\author{Matteo A. C. Rossi}
\email{matteo.rossi@unimi.it}
\homepage{http://users.unimi.it/aqm}
\affiliation{Quantum Technology Lab,
Dipartimento di Fisica, Universit\`a degli Studi di Milano,
20133 Milano, Italy}
\author{Francesco Albarelli}
\email{francesco.albarelli@unimi.it}
\homepage{http://users.unimi.it/aqm}
\affiliation{Quantum Technology Lab,
Dipartimento di Fisica, Universit\`a degli Studi di Milano,
20133 Milano, Italy}
\author{Matteo G. A. Paris}
\email{matteo.paris@fisica.unimi.it}
\homepage{http://users.unimi.it/aqm}
\affiliation{Quantum Technology Lab, Dipartimento di Fisica,
Universit\`a degli Studi di Milano, 20133 Milano, Italy}
\affiliation{CNISM, Unit\`a Milano Statale, I-20133 Milano, Italy}
\affiliation{INFN, Sezione di Milano, I-20133 Milano, Italy}
\date{\today}
\begin{abstract}
We address the characterization of dissipative bosonic channels
and show that estimation of the loss rate by Gaussian probes
(coherent or squeezed) is improved in the presence of Kerr
nonlinearity. In particular, enhancement of precision may be
substantial for short interaction time, i.e. for media of
moderate size, e.g. biological samples. We analyze in detail
the behaviour of the quantum Fisher information (QFI), and
determine the values of nonlinearity maximizing the QFI as
a function of the interaction time and of the parameters of
the input signal. We also discuss the precision achievable
by photon counting and quadrature measurement and present
additional results for truncated, few-photon, probe signals.
Finally, we discuss the origin of the precision enhancement,
showing that it {\em cannot} be linked quantitatively to the
non-Gaussianity of the interacting probe signal.
\end{abstract}
\maketitle
\section{Introduction} 
\label{sec:introduction}
The characterization of quantum channels is a relevant task in quantum
technology
\cite{LoPresti01,Fujiwara2001,LoPresti03,Sarovar2004,Lobino563,Oli07}.
In particular, characterizing lossy channels in continuous variable
systems is crucial to quantify decoherence \cite{Serafini2005}, to
assess quantum illumination protocols
\cite{yuen09,Tan2008,Guha09,Brida10} and to realize quantum reading of
classical memories \cite{Pirandola2011a}. In some specific cases, the
task is simply to discriminate between the presence or the absence of losses
\cite{sasaki97,paris01,Invernizzi2011}, whereas, in general, a strategy
to estimate the exact value of the loss is needed.
\par
The loss rate in optical media and, in turn, the overall loss of the
corresponding channels, are not observable quantities in a strict sense.
As a consequence, one has to infer their value indirectly, i.e. by
assessing the influence of loss on a given probing signal by measuring a
suitably chosen observable.  The overall choice of the probe, of the
measurement, and of the data processing is usually referred to as an
estimation strategy.  Optimization of the estimation strategy, i.e.
minimization of intrinsic and extrinsic fluctuations of the estimate,
may be pursued upon employing quantum estimation theory
\cite{HelstromBook,Braunstein94,Paris2009,Escher2011}, which provides
constructive tools to determine the  initial state of the quantum probe
and the optimal measurement to be performed at the output.  The ultimate
bound on precision is set by the quantum Cramèr-Rao inequality, 
written in terms of the so-called quantum Fisher information.
\par
In the last decades, much attention has been devoted to the estimation
of loss with different initial preparations of the probes. Optimization
over Gaussian input states has been performed \cite{Monras2007},
showing that ultimate precision may be achieved using photon
counting and Gaussian operations at the output. Fock states have also
been shown to saturate the ultimate bound on precision
\cite{Sarovar2004,Adesso2009}, whereas the performances of thermal 
states have been recently investigated \cite{Gea16}. 
The general scenario of lossy
media probed by Gaussian signals at finite temperature has been
considered \cite{Monras2011a}, showing that a two-mode
squeezed vacuum state is optimal for estimating both the loss
parameter and the thermal noise.
The benefit of using entanglement in a specific interferometric
setup has also been discussed \cite{Venzl2007}. Recently the problem of
estimating both the loss and the phase shift in interferometry has
been addressed \cite{Crowley2014}, as well as the related problem
of estimating the efficiency of realistic detectors 
\cite{Barbieri2015,Grandi2015}.
\par
So far, attention has been focused on Gaussian lossy channels where
dissipation is due to linear coupling of the a radiation mode to the
environment, modeled as a bath of external oscillators.  On  the other
hand, optical media where light propagates, such as gasses, biological
samples or optical fibers, may be characterized also by a (usually
small) non-linear response to the electromagnetic field. A question thus
arises on whether estimation of linear loss in the presence of
nonlinearity is enhanced, or not, compared to the pure linear case.
Here, we address this question, by considering systems where
besides dissipation due to linear coupling to the environment,
some form of nonlinearity is present. 
In particular, we focus on self-Kerr interaction
\cite{boyd2008nonlinear}, occurring during propagation of radiation 
in a nonlinear medium with non negligible cubic nonlinearity. The 
Kerr effect has been widely studied in quantum optics either at 
zero  \cite{Milburn1986a} or at finite temperature 
\cite{Stobinska2008}, and attracted interest because
it can be employed to generate Schr\"odinger cat-like states
\cite{Yurke1986,Yurke1988,Miranowicz2011,Paris1999,Jeong2004}. Nonlinearity
of optical fibers has been discussed for it negative impact
on the channel capacity \cite{Essiambre2012}, whereas its role as a resource
in the estimation of losses has not been assesed so far.
\par
As a matter of fact, the presence of non-linear effects has
been already recognized as a resource for quantum estimation, since
it allows one to achieve high precision by using robust classical
probe states, instead of
fragile nonclassical states \cite{Luis2004,Rivas2010,Luis2010c}.  In
particular, Kerr-type nonlinearity may be exploited for
estimation of squeezing and displacement of a Gaussian state
\cite{Genoni2009} and to improve Michaelson interferometry \cite{Luis2015}.
\par
In this paper, we analyze in detail estimation of loss in the presence
of Kerr nonlinearity. We focus mostly on estimation strategies based on
Gaussian probes (coherent and squeezed vacuum states), while also
briefly examining the use of few-photon probes, the simplest nontrivial
ones being optical qutrits. Overall, our results indicate that the
presence of Kerr nonlinearity always enhances estimation, improving
precision compared to the pure linear case.
\par
In particular, by focussing attention on the
estimation of the loss rate parameter of the channel rather than the
overall loss (which also includes the interaction time), we
make the time dependence explicit. This is a relevant feature of our
analysis since dissipation and nonlinearity set two different time scales
in the evolution of the probe state.
In this way, we address both regimes of ``short'' and 
``long" interaction times,
showing that i) nonlinearity always improves estimation; ii)
enhancement of precision may be substantial for short interaction
time, i.e. for media of moderate size.
\par
The paper is structured as follows: in Section
\ref{sec:quantum_estimation_theory} we briefly review the main tools of
quantum estimation theory in order to establish the notation. In Section
\ref{sec:model} we present in detail the interaction model we are
dealing with, whereas in Section \ref{sec:absence_of_non_linear_effects}
we discuss the solution of the problem in the absence of
non-linearities. In Sec \ref{sec:solution_in_presence_of_kerr_effect} we
give an approximate, analytic, solution for the estimation problem with
coherent probes, which holds when the Kerr coupling is much smaller
than the loss
parameter, and present a detailed numerical study for the general
case. We also briefly analyze the use of optical qutrit probes and
discuss whether non-Gaussianity plays a role in the estimation procedure.
Section \ref{sec:conclusions} closes the paper with some concluding
remarks.
\section{Quantum estimation theory} 
\label{sec:quantum_estimation_theory}
Here we briefly review local estimation theory and its generalization to
quantum systems \cite{Paris2009}. In an estimation procedure we want to
infer the value of a parameter, say $\gamma$, from the data collected by
$n$ measurements, $\{x_1,\ldots,x_n\}$. We thus build an estimator
$\hat\gamma (\{x_1,\ldots,x_n\})$, that is a function of the outcomes of
the measurements.
The estimated value of the parameter will be characterized by a
statistical error $\delta \gamma$, which is bounded from below by the
Cramèr-Rao inequality \cite{Cramer1946}
\begin{equation}\label{eq:cramer_rao}
	\delta \gamma^2 \geq \frac 1 {N F(\gamma)},
\end{equation}
where $N$ is the size of the sample data and $F(\gamma)$ is the classical Fisher information (FI), defined as
\begin{equation}\label{eq:fisher_information_definition}
	F(\gamma) = \expected{\left(\frac{\partial \ln p(x|\gamma)}{\partial \gamma}\right)^2}.
\end{equation}
In Eq. \eqref{eq:fisher_information_definition} $p(x|\gamma)$ is the
probability that the outcome of a measurement is $x$ when the value of
the parameter is $\gamma$, and $\expected{\cdot}$ is the expected value
over the probability distribution $p(x|\gamma)$.
\par
If the system is quantum, then $p(x|\gamma) = \Tr (\rho_\gamma \Pi_x)$,
where $\rho_\gamma$ is the density operator and $\Pi_x$ is the POVM
operator for the outcome $x$. By introducing the logarithmic symmetric
derivative $L_\gamma$, satisfying $2\partial_\gamma \rho_\gamma =
L_\gamma \rho_\gamma + \rho_\gamma L_\gamma$, we can rewrite Eq.
\eqref{eq:fisher_information_definition} as \begin{equation}
	F(\gamma) = \expected{\frac{\Re[\Tr(\rho_\gamma \Pi_x L_\gamma)]^2}{\Tr(\rho_\gamma \Pi_x)}}.	
\end{equation}
By maximizing $F(\gamma)$ over all possible quantum measurements on the systems we obtain the quantum Fisher information (QFI) $H(\gamma)$, which has the following expression \cite{Paris2009}:
\begin{equation}
	H(\gamma) = \Tr(\rho_\gamma L_\gamma^2).
\end{equation}
We can thus write a quantum version of the Cramèr-Rao bound,
\begin{equation}\label{eq:quantum_cramer_rao}
	\delta\gamma^2 \geq \frac 1 {N H(\gamma)},
\end{equation}
which gives the ultimate precision achievable on the estimation of $\gamma$ with a quantum measurement. The QFI can be calculated explicitly after a diagonalization of the density operator. Upon writing $\rho_\gamma = \sum_n p_n \ket{\psi_n}\bra{\psi_n}$, we get
\begin{equation}\label{eq:qfi}
	H(\gamma) = 2 \sum_{n,m} \frac{|\braket{\psi_m|\partial_\gamma \rho_\gamma|\psi_n}|^2}{p_n+p_m},
\end{equation}
where the sum is carried out over all $n$ and $m$ such that $p_n+p_m\neq 0$. If the state of the quantum system is pure, $\rho_\gamma = \ket{\psi_\gamma}\bra{\psi_\gamma}$, Eq. \eqref{eq:qfi} reduces to
\begin{equation}\label{eq:qfi_pure_state}
	\begin{split}
	H(\gamma) = 4  & \left[ \braket{\partial_\gamma \psi_\gamma| \partial_\gamma\psi_\gamma} + \braket{\partial_\gamma\psi_\gamma|\psi_\gamma}^2 \right. \\
	& \left. + \braket{\psi_\gamma|\partial_\gamma\psi_\gamma}^2 + |\braket{\partial_\gamma\psi_\gamma|\psi_\gamma}|^2\right].
	\end{split}
\end{equation}
\section{The interaction model} 
\label{sec:model}
In this work we consider a lossy bosonic channel with 
a loss rate parameter $\gamma$, which is the quantity 
that we want to estimate, where non-linear Kerr effect with 
coupling $\tilde\lambda$ is present. In the absence of any 
non-linear effect and working in the interaction picture, 
the density operator $\rho$ for a single bosonic mode 
in the channel satisfies a Lindblad master equation of the form
\begin{equation}\label{eq:master_eq_loss}
	\begin{split}
		\frac{d\rho}{dt} & = \frac{\gamma}{2} \mathcal{L}[a]\rho \\
		& = \gamma( a \rho a^\dagger - \frac 12  a^\dagger a \rho - \frac 12 \rho a^\dagger a),
	\end{split}
\end{equation}
where $a$ is the annihilation operator in the Fock space of the bosonic
mode and $\mathcal L$ is the Lindblad operator. This equation can be
obtained, for instance, from the interaction of the bosonic mode with a
bath of harmonic oscillators at zero temperature. The evolution through
a Gaussian lossy channel can also be represented as the interaction of
the input state with a beam splitter \cite{DAriano1994}, i.e. a bilinear
evolution operator $U(\phi) = \exp [ i \phi(a^{\dag}b + a b^{\dag} )]$;
the auxiliary mode $b$ is traced out at the end and it is initially in
its vacuum state. This picture is connected to the master equation
\eqref{eq:master_eq_loss} by the relation $\tan^2 \phi = e^{\gamma t} -
1$; as a matter of fact in previous works \cite{Monras2007,Adesso2009}
the estimation of $\gamma$ was recast as the estimation of $\phi$.
\par
The Kerr interacton is described by a non-linear term in the Hamiltonian of
the system, namely \begin{equation}\label{eq:Kerr_Hamiltonian}
	H_K = \tilde\lambda (a^\dagger a)^2.
\end{equation}
To take into account this effect, 
the master equation in Eq. \eqref{eq:master_eq_loss} now becomes
\begin{equation}\label{eq:Kerr_master_equation}
	\frac{d\rho}{dt} = -i [H_K,\rho] + \frac{\gamma}{2} \mathcal{L}[a]\rho.
\end{equation}
Upon rescaling the quantities with respect to the loss parameter 
$\gamma$ \begin{equation}
	\tau = \gamma t, \qquad \lambda =\tilde \lambda/\gamma,
\end{equation}
we arrive at
\begin{equation}
	\frac{d\rho}{d\tau} = -i \lambda [(a^\dagger a)^2,\rho] + a \rho a^\dagger - \frac 12  a^\dagger a \rho - \frac 12 \rho a^\dagger a,
\end{equation}
which corresponds to the 
following system of equations for the matrix elements of $\rho$:
\begin{equation}\label{eq:Kerr_master_equation_matrix_el}
	\begin{split}
	\frac{d\rho_{p,q}}{d\tau} = & - \left[ i {\lambda} (p^2-q^2) + \frac 12 (p+q)\right]\rho_{p,q} \\ & + \sqrt{(1+p)(1+q)} \rho_{p+1,q+1}.
	\end{split}
\end{equation}
The solution for the $\rho_{p,q}$ can be found 
easily if the initial state is a coherent state, 
$\rho_0 = \ket{\alpha}\bra{\alpha}$. It reads
\begin{equation}\label{eq:matrix_el_exact}
	\begin{split}
	\rho_{p,q}&(\tau) = \frac{\alpha^p\overline\alpha^q}{\sqrt{p!q!}} \\ & \exp\left\{-\frac 12 (p+q) \Delta \tau - |\alpha|^2\left[ 1-\frac{1-e^{-\Delta \tau}}{\Delta} \right]\right\},
	\end{split}
\end{equation}
where $	\Delta = 1 + 2 i  \lambda (p-q) $. 
\par
We will also consider the case of a squeezed vacuum initial state
$\rho_0=|r\rangle \langle r |$, where we restrict to a real squeezing
parameter $r$, so that the squeezing operator reads $S(r) = \exp\left(
\frac{1}{2} r^2 (a^{\dag 2} - a^{2}) \right)$. The explicit analytical
expression of the matrix elements of the solution with this initial
state can be found in Refs. \cite{Milburn1989,Perinova1988}, but the
matrix elements are known also for arbitrary initial states
\cite{Peinova1990,Chaturvedi1991}. Notice 
that for the lossy channel (i.e. a thermal bath at zero
temperature) these analytical expressions of the matrix elements are
suitable for a numerical computation of the values of the relevant
observables. As a matter of fact it is possible to work in a truncated
Hilbert space in the Fock basis, since the loss only drives the system
into smaller subspaces; this would not be possible if we considered both
loss and noise (i.e. a bath with finite temperature).
Notice also that $\rho(\tau)$ is in general a mixed 
state and cannot be diagonalized explicitly, such that an analytic 
expression for the quantum Fisher information is not available.
\par
We start our analysis by reviewing the analytic 
solutions when the Kerr effect is not present (i.e. 
$\lambda = 0$), and then discuss approximate and numerical 
solutions for the general case of $\lambda \neq 0$.
\section{Solution in the absence of non-linear effects} 
\label{sec:absence_of_non_linear_effects}

\begin{figure}[t]
  \centering
    \includegraphics{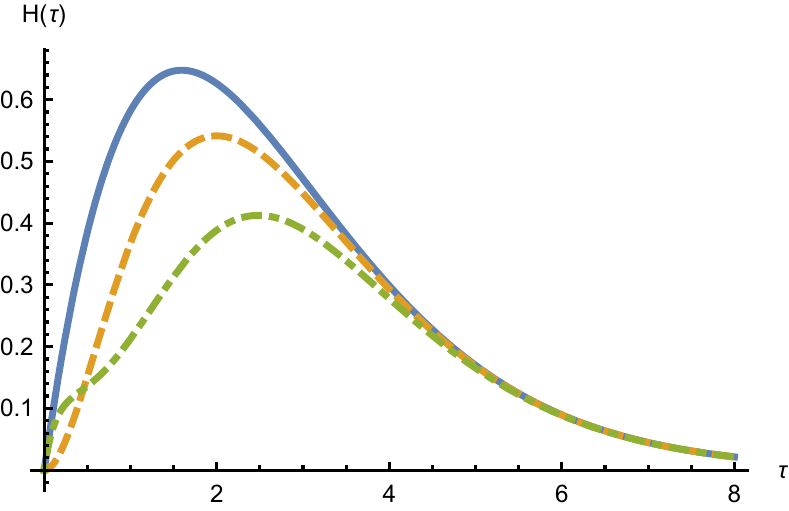}
  \caption{(Color online) Plot of the QFI in the absence of
  non-linearity as a function of the rescaled time $\tau$ for different
  probe states at the fixed mean input energy $\bar{n}=1$. The solid
  blue line represents the optimal Fock state $|1\rangle$, the dashed
  orange line represent a coherent state, while the dot-dashed green
  line represent the squeezed vacuum. The graph reflects the general
  fact that a Fock state is always optimal and for $\tau \to 0$ the
  optimal Gaussian state is the squeezed vacuum, while for greater
  values a coherent state allows for a better estimation.}
  \label{fig:comparison_input_states}
\end{figure}

When $ \lambda = 0$, i.e. the non-linear effects are absent, the channel
is Gaussian and in particular a coherent probe state remains pure and
coherent during the evolution: \begin{equation}\label{eq:state_no_kerr}
	\ket{\psi_\gamma(\tau)} = \ket{\alpha e^{-\frac 12 \tau}}.
\end{equation}
An analytic expression for the QFI is easily obtained using Eq. \eqref{eq:qfi_pure_state}:
\begin{equation}\label{eq:qfi_no_kerr}
	H_\gamma^{\text{c}}(\tau) = \frac{\bar{n}}{\gamma^2} \tau^2 e^{-\tau},
\end{equation}
while for the squeezed vacuum the solution is \cite{Monras2007}:
\begin{equation}\label{eq:qfi_no_kerr_sq}
	H_\gamma^{\text{sv}} (\tau)= \frac{\left(-2 e^{\tau }+e^{2 \tau }+2\right) \tau ^2 \bar{n}}{\gamma ^2 \left(e^{\tau }-1\right) \left(2 e^{\tau } \bar{n}-2 \bar{n}+e^{2 \tau }\right)},
\end{equation}
where $\bar{n}=|\alpha|^2$ for the coherent state and $\bar{n}=\sinh^2 r$ for the squeezed vacuum.
We also report the QFI for Fock probe states $|n\rangle$, which is optimal when the mean energy is an integer ($\bar{n}=n$):
\begin{equation}\label{eq:qfi_no_kerr_fock}
	H_\gamma^{\text{F}}(\tau)=\frac{\bar{n} \tau^2}{\gamma^2 (e^{\tau}-1)}.
\end{equation}
\par
Notice that in general the quantum signal-to-noise ratio (QSNR)
$\gamma^2 H_\gamma(\tau)$ does not depend on $\gamma$: this means that
the bound on the relative error on the  estimation of $\gamma$ is
constant.

\begin{figure}
  \centering
    \includegraphics[width=8cm]{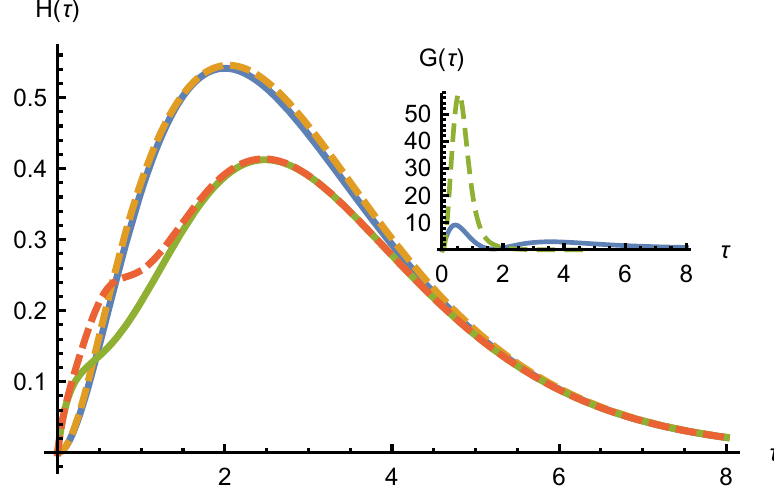}
  \caption{(Color online) Plot of the QFI as a function of the rescaled
  time $\tau$ for different probe states at the fixed mean input energy
  $\bar{n}=1$. The solid curves are obtained in the absence of
  nonlinearity, while the dashed curves are obtained in the presence of Kerr
  nonlinearity (with $\lambda=0.5$).  The solid blue and dashed oranges
  curves which lie on top in the region $\tau \approx 2$ refer to the
  coherent state probe, while the solid green and dashed orange curves
  which lie on top in the region $\tau \approx 0$ refer to the squeezed
  vacuum probe. In the inset panel we represent the relative gain
  $G(\tau) \equiv H_{\lambda,\gamma}(\tau)/H_\gamma(\tau) - 1$ of the
  QFI in the presence of non-linearity over the QFI without Kerr effect,
  shown in percentage. The solid blue line represents the coherent
  probe, while the dashed green line represents the squeezed vacuum. In
  both cases there is a peak in gain at $\tau \lesssim 1$, much more
  pronounced for the squeezed vacuum state. The gain vanishes for
  increasing $\tau$, but a second, smaller peak can be observed for the
  coherent state.} \label{fig:gain_time}
\end{figure}
In Figure \ref{fig:comparison_input_states} we represent the plots of
the QFI for the three probe states; this also sums up previous results
\cite{Monras2007,Adesso2009} by showing that for small losses the
optimal Gaussian state is the squeezed vacuum, for higher losses a
coherent state is better, while a Fock state is optimal for every
$\tau$. Moreover, we observe that in general $H_\gamma(\tau)$ vanishes
for $\tau \gg 1$ and has a global maximum at a certain time $\overline
\tau$. This means that if one is able to control the interaction time in
an experiment, setting it to $\overline\tau$ allows for optimal
estimation of $\gamma$.
In particular for the coherent state the optimal time is $\overline \tau = 2$, with the following optimal value:
\begin{equation}\label{eq:qfi_no_kerr_opt}
	\overline H_\gamma^{\text{c}} = \frac{4 |\alpha|^2}{e^2 \gamma ^2}.
\end{equation}
As a matter of fact, for coherent states the QFI is saturated by
photon-number and a quadrature measurement. Let us compute the
Fisher information (FI) for these two measurements:
The probability distribution for a photon counting experiment for the state is a Poisson distribution with mean $\mu = |\alpha e^{-\tau/2}|^2$. The FI for a Poissonian is $\mu^{-1}$, hence, using the chain rule of derivatives, we get
\begin{equation}
	\begin{split}
		F_n(\gamma,\tau) & = \left(\frac{\partial \tau}{\partial \gamma}\right)^2 \left(\frac{\partial \mu}{\partial \tau}\right)^2 \frac 1 \mu =\frac{|\alpha|^2}{\gamma^2} \tau^2e^{-\tau}.
	\end{split}
\end{equation}
The probability distribution for the quadrature measurement $x= (a + a^\dagger) / \sqrt{2}$ is
\begin{equation}
	p(x|\gamma) = |\braket{x|\alpha e^{-\tau / 2}}|^2 = \frac{e^{-\left(x-\sqrt{2} \Re(\alpha) e^{-\tau/2}\right)^2}}{\sqrt{\pi }}
\end{equation}
and hence the Fisher information, Eq. \eqref{eq:fisher_information_definition}, is
\begin{equation}
	F_x(\gamma,t) = \frac{\tau^2 e^{-\tau}\Re(\alpha)^2}{\gamma^2}.
\end{equation}
We see that $F_x(\gamma,t)= H_\gamma(t)$ as long as $\alpha$ is chosen to be real. If $\alpha$ has a complex phase it suffices to choose the proper quadrature or to apply a phase shift to the coherent state to saturate the QFI.

\section{Solution in the presence of Kerr effect}
\label{sec:solution_in_presence_of_kerr_effect}

As stated in Section \ref{sec:model}, with $\lambda \neq 0$ the state $\rho(\tau)$ is a mixed state and not explicitly diagonalizable. In the following, we present an approximate solution for the coherent probe state, valid in the regime of small $\lambda$ and $\tau$, in which the state of the system remains pure and it is thus possible to get an analytical expression for the QFI. Then we show numerical results obtained from a truncation of the Fock space for both coherent and squeezed vacuum probe states. The results are presented both for the optimal time and small time cases; at optimal time only the coherent input is considered since the optimal value of the QFI is always greater than the optimal value of the squeezed vacuum QFI. This fact can be seen in Fig. \ref{fig:gain_time}, where we show the behavior of the QFI with and without Kerr interaction for both the Gaussian probes we are considering. From the particular choice of parameters in Fig. \ref{fig:gain_time} we see that the QFI with nonlinear interaction always has a greater value: we will show that this is true in general.

\subsection{Pure state approximation}
\label{sub:pure_state_approximation}

\begin{figure}[t]
	\includegraphics{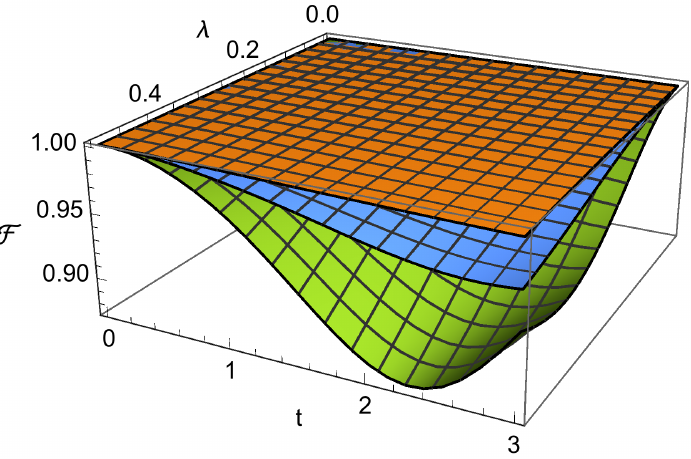}
	\caption{(Color online) Fidelity between the pure state of Eq.
	\eqref{eq:pure_state_approx} and the exact state (truncation at
	10 photons), for $\alpha = 0.5$ (orange), $\alpha = 0.75$ (blue)
	and $\alpha = 1$ (green). The fidelity decreases with increasing
	$\lambda$ and $\alpha$. It temporarily decreases with time, but
	it tends asymptotically to one as the system reaches the state
	$\ket{0}$. For small values of $\alpha$ and $\lambda$ the pure
	state approximation has fidelity above 0.99, which then
	decreases as the energy of the state increases.}
\label{fig:fidelity}
\end{figure}
When we work with a coherent input state and the non-linear effect is
small compared to the loss parameter, i.e. when $ \lambda \ll 1$, the
state of the system can still be approximated with a pure state for
small $\tau$. Expansion of the exponent of $e$ in Eq.
\eqref{eq:matrix_el_exact} to the first order in $ \lambda$ and then
expansion to the second order of $\tau$ yields
\begin{equation}\label{eq:pure_state_approx}
	\begin{split}
	\rho_{p,q}(\tau) = & \frac{\alpha^p\overline\alpha^q}{\sqrt{p!q!}}  \exp\left\{-\frac 12 (p+q)  \tau - e^{-\tau} |\alpha|^2 \right. \\
	& \left.- i  \lambda (p^2 - q^2) \tau - i  \lambda |\alpha|^2 (p - q)\tau^2  \right\}.
	\end{split}
\end{equation}
This is the lowest order of expansion for which we obtain a correction to the quantum Fisher information of Eq. \eqref{eq:qfi_no_kerr}.

The QFI computed for $\rho_{p,q}(\tau)$ of Eq. \eqref{eq:pure_state_approx} is
\begin{equation}\label{eq:qfi_pure_state_approx}
	H_{\lambda,\gamma}^{\text{c}}(\tau) = \frac {\left| \alpha \right| ^2}{\gamma^2} \tau^2 e^{- \tau} \left(1+4 \lambda ^2 \tau^2 \left| \alpha \right| ^4\right) + O(\lambda^3).
\end{equation}

We notice that $H_{\lambda,\gamma}^{\text{c}}(t)$ adds a correction of second order in $\lambda$ and in $\tau$ to $H_\gamma^{\text{c}}(\tau)$ of Eq. \eqref{eq:qfi_no_kerr}.
If we define the \emph{relative gain} in the estimation of $\gamma$ as $G_{\lambda}(\tau)\equiv H_{\lambda,\gamma}(\tau)/H_\gamma(\tau) - 1$, then using the pure state approximation it reads:
\begin{equation}\label{eq:relgain_coh}
	G_{\lambda}^{\text{c}}(\tau)=4\lambda^2 \tau^2 |\alpha|^4 + O(\lambda^3).
\end{equation}

The optimal time, up to the second order in $\lambda$, is
\begin{equation}
	\overline{\tau}(\lambda) = 2 + 32  \lambda^2 |\alpha|^4 + O(\lambda^3)
\end{equation}
and the corresponding optimal QFI is
\begin{equation}\label{eq:qfi_opt_pure_state_approx}
	\overline{H}_\gamma^{\text{c}}(\lambda) = \frac{4 \left| \alpha \right| ^2}{e^2\gamma^2}(1 + 16  \lambda ^2 \left| \alpha \right| ^4)+O(\lambda^3);
\end{equation}
so the \emph{optimal relative gain}  $\overline{G}_{\lambda} \equiv \overline{H}_{\lambda,\gamma}/\overline{H}_\gamma - 1$ is
\begin{equation}\label{eq:relgain_opt_coh}
	\overline{G}_{\lambda}^{\text{c}}=16 \lambda^2 |\alpha|^4 + O(\lambda^3).
\end{equation}

Equations \eqref{eq:relgain_coh} and \eqref{eq:relgain_opt_coh} show that the correction to the QFI due to the presence of a small non-linear effect is positive and increases with $\lambda^2$. This means that the nonlinearity of the dispersive medium can be a resource in the estimation of the loss parameter.

The fidelity of the approximate state of Eq. \eqref{eq:pure_state_approx} to the exact state (after a truncation of the density matrix) is shown in Fig. \ref{fig:fidelity} as a function of $\tau$ and $\lambda$, for two values of $|\alpha|$. The pure state approximation is good for a wide range of parameters only if the energy of the initial state is not too big, so that fidelity is close to one \cite{Bina2014,Mandarino2014}. This means that the analytical expression of the optimal relative gain \eqref{eq:relgain_opt_coh} is good only for small energies, while at a fixed small time $\tau \ll 1$ the relative gain \eqref{eq:relgain_coh} is a good approximation even for higher input energies.

In Subsection \ref{sub:numerical_results} we calculate the QFI numerically for general values of $\lambda$ and $\alpha$, in order to verify the increase of the QFI also for regions where the pure-state approximation does not hold.

\subsection{Numerical results}
\label{sub:numerical_results}
As the density matrix cannot be diagonalized in general and the Fock
space is infinite-dimensional, in order to evaluate the QFI we resort to
numerical diagonalization of the density matrix in a truncated Fock
space. The truncation size, which depends on the input energy, is chosen
in such a way that the difference between the analytical and the
numerical QFI for $\lambda=0$ must be less than $0.001\%$.

\begin{figure}
\includegraphics{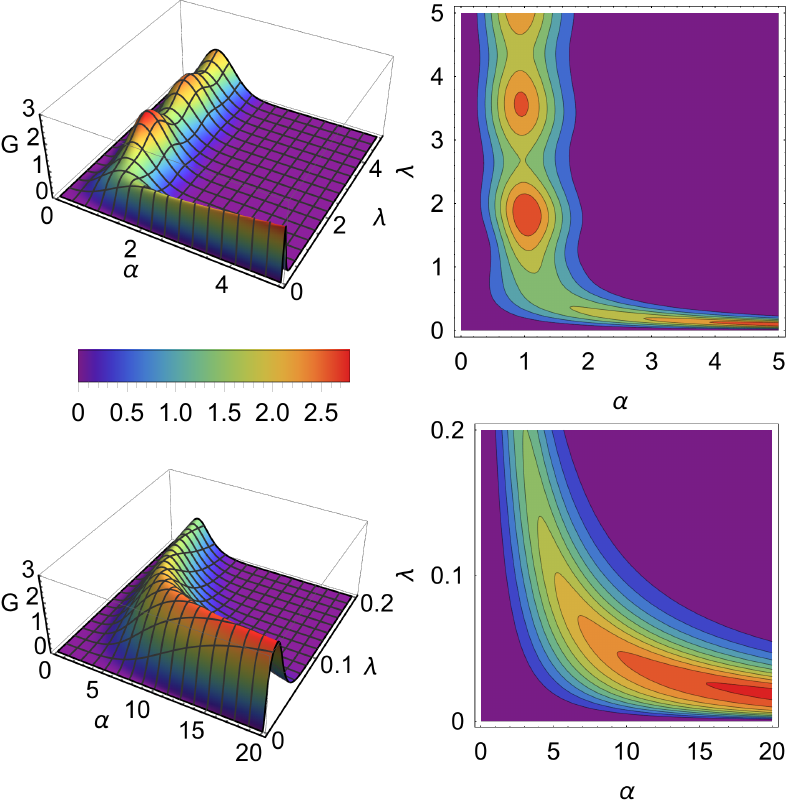}
\caption{(Color online) Optimal relative gain $\overline{G}\equiv
\overline{H}_{\lambda,\gamma}/\overline{H}_\gamma - 1$ of the optimal
QFI in presence of non-linearity over the optimal QFI without Kerr
effect for different regions of $\alpha$ and $\lambda$, shown in
percentage. On the left, a 3D plot, on the right the corresponding
contour plot. We can see that the gain is always greater than zero,
vanishing for large $\lambda$ and $\alpha$. We can identify two regimes:
The first regime, visible in the upper panels when $\alpha \lesssim 2$
is characterized by the presence of local maxima of the gain, which
reaches values of about $2 \%$. For large $\lambda$ the improvement
reaches a non-vanishing asymptotical value. In the second regime,
visible in the lower panels, at fixed $\alpha$ the gain has a single
maximum with respect to $\lambda$. As $\alpha$ increases, the maximum
moves to smaller values of $\lambda$, but $G$ increases.}
\label{fig:improvement}
\end{figure}

\begin{figure*}[t]
  \centering
\includegraphics[width=.3\textwidth]{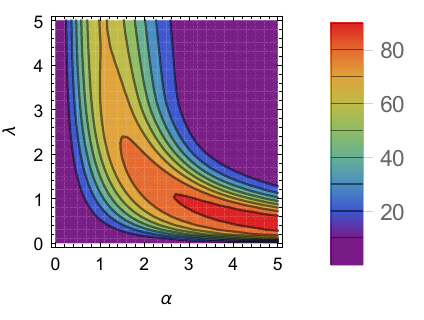}
\includegraphics[width=.3\textwidth]{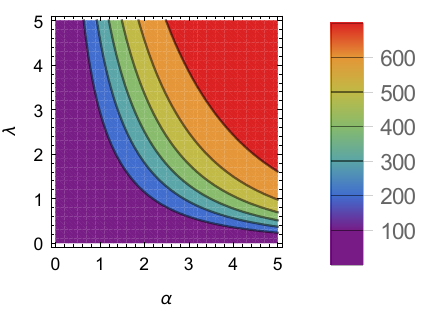}
\includegraphics[width=.3\textwidth]{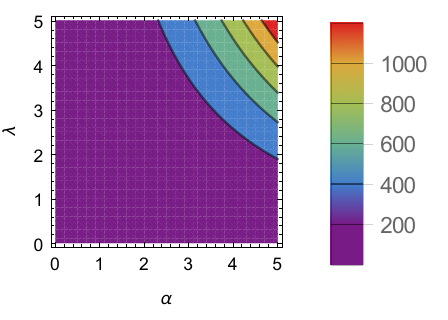}
\includegraphics[width=.3\textwidth]{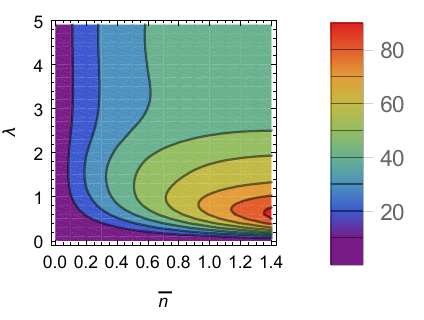}
\includegraphics[width=.3\textwidth]{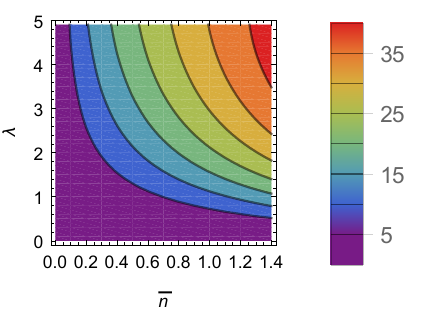}
\includegraphics[width=.3\textwidth]{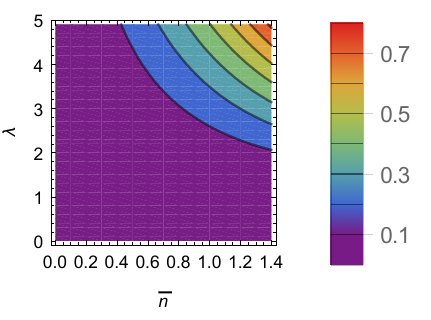}
  \caption{(Color online) Relative gain $G(\tau) \equiv H_{\lambda,\gamma}(\tau)/H_\gamma(\tau) - 1$ of the QFI in presence of non-linearity over the QFI without Kerr effect at fixed time for a coherent probe state (top) and for a squeezed vacuum probe state (bottom), shown in percentage. From left to right we have the results for $\tau=0.5$, $0.1$, $0.01$. For coherent states we can see a structure similar to that of Fig. \ref{fig:improvement}: the relative gain increases with $\alpha$ and $\lambda$ until it reaches a maximal value, but at small $\tau$ the relative gain is much higher than at the optimal time. For the squeezed vacuum state the gain is smaller as $\tau$ gets smaller (cfr. Fig. \ref{fig:gain_time}).}
  \label{fig:improvement_smallt}
\end{figure*}

\subsubsection{Optimal QFI}
The behavior of the QFI as a function of time for fixed $\lambda$ and $\alpha$ is shown in Fig. \ref{fig:gain_time}. The QFI starts from zero and reaches a maximum, then vanishes as $\tau$ increases and the system reaches the zero-photon state $\ket 0$. Assuming that we are able to control the interaction time of the probe with the channel, we can consider as a figure of merit the optimal QFI, i.e. the maximum of $H_{\lambda,\gamma}(t)$ over time.

In Fig. \ref{fig:improvement} we show the optimal relative gain in the estimation of $\gamma$. The first notable result is the confirmation of the results obtained in the pure state approximation: the optimal QFI in presence of non-linearity is always greater than without Kerr effect, i.e. the optimal relative gain is always greater than zero. It vanishes for increasing $\alpha$ and $\lambda$ and for $\alpha \rightarrow 0$.

By looking at the panels of Fig. \ref{fig:improvement}, we can identify two regimes. The first regime, for $\alpha \lesssim 2$, is characterized by the presence of local maxima of the gain. At fixed $\alpha$, the maxima occur periodically, with $G$ reaching an asymptotic value for $\lambda \rightarrow \infty$. In the second regime, for $\alpha \gtrsim 2$, there is a single local maximum for the gain at fixed $\alpha$. For increasing $\alpha$, the optimal $\lambda$ decreases, but $G$ increases. It is not clear if there is a local maximum for $\alpha$ greater than the values under investigation or if this behavior will persist for $\alpha \rightarrow \infty$, and, in the latter case, if $G$ increases indefinitely or saturates with $\alpha$.

\subsubsection{Small time QFI}

Now instead of studying the QFI maximized over time we look at the behavior at a fixed time, in particular we focus on times smaller than the characteristic time of the loss, i.e. $\tau < 1$, as an example we study three cases $\tau=0.5,0.1,0.01$. This regime is of interest for media of moderate size, such as biological samples.

In this setting the improvement brought by the nonlinear interaction can be substantial. In Fig. \ref{fig:improvement_smallt} we show the results for a coherent probe state (top row) and for a squeezed vacuum probe state (bottom row). For the squeezed probe we restricted the computation to a smaller range of mean input energies, as the dimension of the truncated Hilbert space needed to obtain a good approximation grows much more rapidly.

By looking at the top-left panel in Fig. \ref{fig:improvement_smallt}, the one for $\tau=0.5$, we notice a similar structure to the one in Fig. \ref{fig:improvement}, albeit rescaled. We found that fixing the time parameter $\tau$ changes the scaling in the $\alpha-\lambda$ (or $\bar{n}-\lambda$) plane; however, it was not possible to explicitly see this scaling from the analytical expressions of the states.

The improvement due to the Kerr nonlinearity is much more relevant at times which do not correspond to the optimal time, indeed in Fig. \ref{fig:gain_time} we see that the maxima of the graph in the inset panel do not correspond to the ones in the main graph. Moreover, even if the behavior of different input states is slightly different, the most relevant improvement is always obtained for $\tau<1$, this is due to the fact that the value of the QFI at those times is smaller, so that a slight improvement in the absolute value brings a great relative gain.

\subsubsection{FI for the quadrature measurement with coherent probe} 
\label{ssub:fi_for_the_quadrature_measurement}

Although the optimal QFI is improved by the Kerr effect, we need to find the actual measurement that reaches the quantum bound.
In Section \ref{sec:absence_of_non_linear_effects} we showed that for a coherent probe both photon counting and quadrature measurement are optimal when $\lambda = 0$, however they are not optimal if the nonlinear term is present.
Indeed, photon counting is not affected at all by the Kerr effect, as the diagonal elements of the density matrix are independent of $\lambda$. For this reason we study numerically the effect of nonlinearity on a quadrature measurement.
We present the results for a coherent probe state; the analysis is less interesting for a squeezed vacuum probe as the optimal measurement in the linear case is not just a quadrature measurement, but is given by Gaussian operations and photon counting \cite{Monras2007}.

\begin{figure}[t]
	\includegraphics{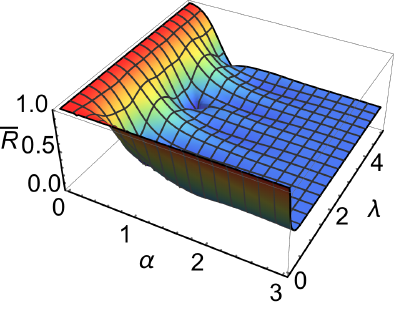}
	\includegraphics{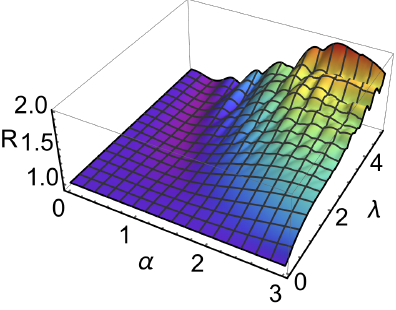}
	\caption{(Color online) In the left panel we show the ratio $\overline R =
	F_x(\overline \tau)/\overline H_\gamma^c(\lambda)$ between the 
	FI of the
	quadrature at the time $\overline\tau$, $F_x(\overline \tau)$,
	after an optimization over the quadrature phase
	and the
	optimal QFI $\overline H_\gamma(\lambda)$,
	for various
	values of $\lambda$ and $\alpha$. The quadrature measurement is
	optimal only for $\lambda = 0$ and for vanishing energy of the
	probe ($\alpha \rightarrow 0$). For $\alpha \lesssim 2$ the
	ratio oscillates with $\lambda$. For large $\alpha$ and
	$\lambda$ the ratio reaches asymptotically the value of $1/3$.
	In the right panel we show the ratio
	$R=F_x(\tau)/H_{\gamma}^{c}(\tau)$ for fixed small $\tau=0.1$;
	the quantity $H_{\gamma}^c$ is the QFI without nonlinearities
	(Eq. \eqref{eq:qfi_no_kerr}). The quadrature measurement in
	presence on Kerr effect achieves increasingly better
	performances for increasing values of $\lambda$ and $\alpha$,
	even if the ratio has a slightly oscillating behaviour and there are some regions in which $R < 1$, i.e. the Kerr effect is slightly detrimental.}
	\label{fig:FI_vs_QFI}
\end{figure}

We found that in general the quadrature measurement is not optimal, i.e. the Fisher information is always lower than the QFI. This fact is presented in the left panel of Fig. \ref{fig:FI_vs_QFI}, for measurements at the optimal time, where the ratio $\overline R = F_x(\overline \tau)/\overline H_\gamma(\lambda)$ is shown. Here $\overline H_\gamma(\lambda)$ is the optimal QFI and $F_x(\overline \tau)$ is the FI of the quadrature measurement at the time $\overline\tau$ that optimizes the QFI, after an optimization over the quadrature phase (the optimal quadrature phase depends on $\alpha$ and $\lambda$). The ratio is close to one only for $\lambda$ close to zero or $\alpha \ll 1$. For increasing $\alpha$ and $\lambda$ the ratio appears to tend asymptotically to $1/3$.

In the small time regime a quadrature measurement is still sub-optimal in presence of nonlinearity, however in some cases such a measurement can perform better than the best possible measurement in the linear case, because the relative improvement of the QFI in this regime is substantial.

In particular, this behaviour seems to increase with increasing nonlinearity $\lambda$ and increasing input energy $\alpha$, however we can see from the right panel of Fig. \ref{fig:FI_vs_QFI} that oscillations are present and there are small regions where a quadrature measurement does not give an improvement, i.e. $R < 1$.

\subsection{Results with optical qutrit states}
\label{subsec:qutrit}

One may wonder what happens if the optimal Fock states are used as probes, instead of Gaussian states.

The obvious answer is that the Kerr nonlinear term $(a^{\dag} a)^2$ does not affect single Fock states, but also a simple superpositions of the form $a |0\rangle + b |n \rangle$ is not affected. The most simple superposition affected by the nonlinear evolution is the optical qutrit state
\begin{equation}
\label{eq:qutrit}
\cos\theta |0\rangle +   e^{i \mu }\sin\theta\sin\varphi
|1\rangle  + e^{i \nu }\sin\theta \cos\varphi
 |2 \rangle,
\end{equation}
where $\theta$ is fixed by choosing the mean energy $\bar{n}$ as the relevant parameter, so that $\theta = \arcsin \sqrt{2 \bar{n}/(3 + \cos 2 \varphi)}$.

In the Gaussian lossy evolution, without Kerr nonlinearity, these qutrit states approximate the optimal non-Gaussian states when the mean energy $\bar{n}$ is not an integer; this is particularly important for the low energy regime $\bar{n}<1$ \cite{Adesso2009}.

In general, the maximum value of the QFI obtainable with the state \eqref{eq:qutrit} is the same regardless of the Kerr term in the evolution, but the maximum happens for different values of the initial parameters and at a different time. This is due to the fact that during the evolution the system is constrained to remain in the subspace of dimension three; so if we optimize on every possible parameter there is no room for improvement left.

However in order to achieve the maximal QFI one should be able to tune the value of the initial parameters for every mean energy $\bar{n}$, and in the nonlinear case also for every value of $\lambda$. In particular in the linear case the result must be optimized only over the parameter $\varphi$, since the relative phases $\mu$ and $\nu$ give an optimal result for the value $\pi$.

\begin{figure}
	\includegraphics{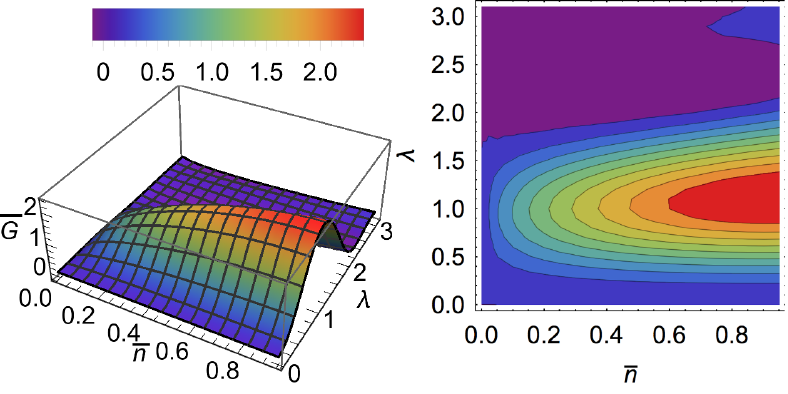}
	\caption{(Color online) Average relative gain of the optimal QFI in presence of non-linearity over the optimal QFI without Kerr effect for qutrit states, shown in percentage. The range of the parameters are $0<\bar{n}<1$ and $0<\lambda<\pi$. On the left, a 3D plot, on the right the corresponding contour plot. Every point in the plot is the average improvement obtained by generating 1000 random values of the parameter $\varphi$ of the state \eqref{eq:qutrit} in the range $\left(0,\frac{\pi}{2}\right)$, while the phases are fixed $\mu=\nu=\pi$ and $\theta$ is fixed by the choice of the mean energy $\bar{n}$.}
	\label{fig:improvement_qutrit}
\end{figure}

We thus resort to work in a setting similar to the one used to study the optimal gain for the coherent states: given a \emph{fixed} initial state we check if the nonlinear evolution brings an improvement. In particular we fix $\mu=\nu=\pi$ and we check the behaviour of the quantum Fisher information for different values of $\varphi$, while optimizing over time $t$. The results are in Fig. \ref{fig:improvement_qutrit}: we find that on average the nonlinear terms brings an improvement for values of $\lambda \approx 1$, i.e. when the nonlinear parameter is approximately equal to the loss parameter to estimate. For higher values of $\lambda$ we have an oscillatory behaviour and on average the nonlinearity can also be detrimental.

We also found that at fixed small times the nonlinear Kerr term does not always bring an improvement on average when using qutrit states.

\subsection{Discussion}
The nonlinear Kerr interaction makes the
initial Gaussian probe non-Gaussian during the evolution
and a question arises on whether the observed increase of the
QFI may be quantitatively linked to
some quantifier of non-Gaussianity \cite{Genoni2010}.
Indeed, it would be desirable to identify the proper resource which guarantees
the improvement in the estimation by means of a nonlinear interaction,
since this would represent a guideline to engineer optimal estimation schemes.
On the other hand also a qualitative indicator to assess the
effectiveness of Kerr interaction to enhance precision may be useful.
\par
In previous works it has been conjectured \cite{Adesso2009} that a
family of optimal non-Gaussian states exists for any fixed energy, but
the authors remark that non-Gaussianity in itself cannot be a resource
since there are non-Gaussian states which are far less efficient probes
than the optimal Gaussian ones. Hereby we confirm that result. In fact,
during its evolution a Gaussian input state first becomes non-Gaussian
and then it evolves towards the Gaussian state $\ket{0}$, which is
the stationary state. This qualitative behaviour is
also shown by the relative gain in the estimation of $\gamma$, as can be seen in
Fig. \ref{fig:gain_time}. These two quantities, however, do not have a
quantitative relation in general, e.g. states leading to the largest
improvement at optimal time are not the most non-Gaussian.
\par
Overall, our results show that while the evolution drives the Gaussian
input into a set of non-Gaussian states which are more sensitive to
loss detection, non-Gaussianity is not a resource in itself. This
idea is confirmed by looking at the behaviour of qutrit probe states,
which are already highly non-Gaussian: there we find evidences that 
the Kerr interaction may be detrimental in some regimes, whereas 
when an improvement is present, the states are non necessarily more 
non-Gaussian.
\section{Conclusions} 
\label{sec:conclusions}
In conclusion, we have addressed the characterization of 
dissipative bosonic channels in the presence of nonlinearity
and shown that the estimation of the loss rate by coherent or 
squeezed probes is improved in the presence of Kerr
nonlinearity. 
In particular, enhancement of precision may be
substantial for short interaction time, i.e. for media of
moderate size, whereas for larger media the improvement
is asymptotically negligible.
\par
We have analyzed in detail
the behaviour of the quantum Fisher information (QFI), and 
have found the values of nonlinearity maximizing the QFI as
a function of the interaction time and of the parameters of
the input signal. We have also shown that Ker nonlinearity 
may be helpful also using few photon probes as optical
qutrits.
\par
We have discussed the precision achievable
by photon counting and quadrature measurement, showing that
they cannot, in general, achieve the QFI in the presence
of nonlinearity. On the other hand, for short interaction times
even this suboptimal measurement offers a precision improvement 
compared to the linear case.  
\par
Finally, we have discussed the possible origin of the precision enhancement,
showing that it cannot be linked quantitatively to the
non-Gaussianity of the interacting probe signal.
\begin{acknowledgments}
The authors thank Benoit Vallet for his contribution in the early stage
of this project. This work has been supported by EU through the
Collaborative Project QuProCS (Grant Agreement 641277) and by UniMI
through the H2020 Transition Grant 15-6-3008000-625.
\end{acknowledgments}
\bibliography{libLK}

\begin{thebibliography}{50}%
\makeatletter
\providecommand \@ifxundefined [1]{%
 \@ifx{#1\undefined}
}%
\providecommand \@ifnum [1]{%
 \ifnum #1\expandafter \@firstoftwo
 \else \expandafter \@secondoftwo
 \fi
}%
\providecommand \@ifx [1]{%
 \ifx #1\expandafter \@firstoftwo
 \else \expandafter \@secondoftwo
 \fi
}%
\providecommand \natexlab [1]{#1}%
\providecommand \enquote  [1]{``#1''}%
\providecommand \bibnamefont  [1]{#1}%
\providecommand \bibfnamefont [1]{#1}%
\providecommand \citenamefont [1]{#1}%
\providecommand \href@noop [0]{\@secondoftwo}%
\providecommand \href [0]{\begingroup \@sanitize@url \@href}%
\providecommand \@href[1]{\@@startlink{#1}\@@href}%
\providecommand \@@href[1]{\endgroup#1\@@endlink}%
\providecommand \@sanitize@url [0]{\catcode `\\12\catcode `\$12\catcode
  `\&12\catcode `\#12\catcode `\^12\catcode `\_12\catcode `\%12\relax}%
\providecommand \@@startlink[1]{}%
\providecommand \@@endlink[0]{}%
\providecommand \url  [0]{\begingroup\@sanitize@url \@url }%
\providecommand \@url [1]{\endgroup\@href {#1}{\urlprefix }}%
\providecommand \urlprefix  [0]{URL }%
\providecommand \Eprint [0]{\href }%
\providecommand \doibase [0]{http://dx.doi.org/}%
\providecommand \selectlanguage [0]{\@gobble}%
\providecommand \bibinfo  [0]{\@secondoftwo}%
\providecommand \bibfield  [0]{\@secondoftwo}%
\providecommand \translation [1]{[#1]}%
\providecommand \BibitemOpen [0]{}%
\providecommand \bibitemStop [0]{}%
\providecommand \bibitemNoStop [0]{.\EOS\space}%
\providecommand \EOS [0]{\spacefactor3000\relax}%
\providecommand \BibitemShut  [1]{\csname bibitem#1\endcsname}%
\let\auto@bib@innerbib\@empty
\bibitem [{\citenamefont {D'Ariano}\ and\ \citenamefont
  {Lo~Presti}(2001)}]{LoPresti01}%
  \BibitemOpen
  \bibfield  {author} {\bibinfo {author} {\bibfnamefont {G.~M.}\ \bibnamefont
  {D'Ariano}}\ and\ \bibinfo {author} {\bibfnamefont {P.}~\bibnamefont
  {Lo~Presti}},\ }\href {\doibase 10.1103/PhysRevLett.86.4195} {\bibfield
  {journal} {\bibinfo  {journal} {Phys. Rev. Lett.}\ }\textbf {\bibinfo
  {volume} {86}},\ \bibinfo {pages} {4195} (\bibinfo {year}
  {2001})}\BibitemShut {NoStop}%
\bibitem [{\citenamefont {Fujiwara}(2001)}]{Fujiwara2001}%
  \BibitemOpen
  \bibfield  {author} {\bibinfo {author} {\bibfnamefont {A.}~\bibnamefont
  {Fujiwara}},\ }\href {\doibase 10.1103/PhysRevA.63.042304} {\bibfield
  {journal} {\bibinfo  {journal} {Phys. Rev. A}\ }\textbf {\bibinfo {volume}
  {63}},\ \bibinfo {pages} {042304} (\bibinfo {year} {2001})}\BibitemShut
  {NoStop}%
\bibitem [{\citenamefont {D'Ariano}\ and\ \citenamefont
  {Lo~Presti}(2003)}]{LoPresti03}%
  \BibitemOpen
  \bibfield  {author} {\bibinfo {author} {\bibfnamefont {G.~M.}\ \bibnamefont
  {D'Ariano}}\ and\ \bibinfo {author} {\bibfnamefont {P.}~\bibnamefont
  {Lo~Presti}},\ }\href {\doibase 10.1103/PhysRevLett.91.047902} {\bibfield
  {journal} {\bibinfo  {journal} {Phys. Rev. Lett.}\ }\textbf {\bibinfo
  {volume} {91}},\ \bibinfo {pages} {047902} (\bibinfo {year}
  {2003})}\BibitemShut {NoStop}%
\bibitem [{\citenamefont {Sarovar}\ and\ \citenamefont
  {Milburn}(2006)}]{Sarovar2004}%
  \BibitemOpen
  \bibfield  {author} {\bibinfo {author} {\bibfnamefont {M.}~\bibnamefont
  {Sarovar}}\ and\ \bibinfo {author} {\bibfnamefont {G.~J.}\ \bibnamefont
  {Milburn}},\ }\href {\doibase 10.1088/0305-4470/39/26/015} {\bibfield
  {journal} {\bibinfo  {journal} {J. Phys. A}\ }\textbf {\bibinfo {volume}
  {39}},\ \bibinfo {pages} {8487} (\bibinfo {year} {2006})}\BibitemShut
  {NoStop}%
\bibitem [{\citenamefont {Lobino}\ \emph {et~al.}(2008)\citenamefont {Lobino},
  \citenamefont {Korystov}, \citenamefont {Kupchak}, \citenamefont {Figueroa},
  \citenamefont {Sanders},\ and\ \citenamefont {Lvovsky}}]{Lobino563}%
  \BibitemOpen
  \bibfield  {author} {\bibinfo {author} {\bibfnamefont {M.}~\bibnamefont
  {Lobino}}, \bibinfo {author} {\bibfnamefont {D.}~\bibnamefont {Korystov}},
  \bibinfo {author} {\bibfnamefont {C.}~\bibnamefont {Kupchak}}, \bibinfo
  {author} {\bibfnamefont {E.}~\bibnamefont {Figueroa}}, \bibinfo {author}
  {\bibfnamefont {B.~C.}\ \bibnamefont {Sanders}}, \ and\ \bibinfo {author}
  {\bibfnamefont {A.~I.}\ \bibnamefont {Lvovsky}},\ }\href {\doibase
  10.1126/science.1162086} {\bibfield  {journal} {\bibinfo  {journal}
  {Science}\ }\textbf {\bibinfo {volume} {322}},\ \bibinfo {pages} {563}
  (\bibinfo {year} {2008})}\BibitemShut {NoStop}%
\bibitem [{\citenamefont {Olivares}\ and\ \citenamefont {Paris}(2007)}]{Oli07}%
  \BibitemOpen
  \bibfield  {author} {\bibinfo {author} {\bibfnamefont {S.}~\bibnamefont
  {Olivares}}\ and\ \bibinfo {author} {\bibfnamefont {M.~G.~A.}\ \bibnamefont
  {Paris}},\ }\href {\doibase 10.1103/PhysRevA.76.042120} {\bibfield  {journal}
  {\bibinfo  {journal} {Phys. Rev. A}\ }\textbf {\bibinfo {volume} {76}},\
  \bibinfo {pages} {042120} (\bibinfo {year} {2007})}\BibitemShut {NoStop}%
\bibitem [{\citenamefont {Serafini}\ \emph {et~al.}(2005)\citenamefont
  {Serafini}, \citenamefont {Paris}, \citenamefont {Illuminati},\ and\
  \citenamefont {Siena}}]{Serafini2005}%
  \BibitemOpen
  \bibfield  {author} {\bibinfo {author} {\bibfnamefont {A.}~\bibnamefont
  {Serafini}}, \bibinfo {author} {\bibfnamefont {M.~G.~A.}\ \bibnamefont
  {Paris}}, \bibinfo {author} {\bibfnamefont {F.}~\bibnamefont {Illuminati}}, \
  and\ \bibinfo {author} {\bibfnamefont {S.~D.}\ \bibnamefont {Siena}},\ }\href
  {\doibase 10.1088/1464-4266/7/4/R01} {\bibfield  {journal} {\bibinfo
  {journal} {J. Opt. B}\ }\textbf {\bibinfo {volume} {7}},\ \bibinfo {pages}
  {R19} (\bibinfo {year} {2005})}\BibitemShut {NoStop}%
\bibitem [{\citenamefont {Yuen}\ and\ \citenamefont {Nair}(2009)}]{yuen09}%
  \BibitemOpen
  \bibfield  {author} {\bibinfo {author} {\bibfnamefont {H.~P.}\ \bibnamefont
  {Yuen}}\ and\ \bibinfo {author} {\bibfnamefont {R.}~\bibnamefont {Nair}},\
  }\href {\doibase 10.1103/PhysRevA.80.023816} {\bibfield  {journal} {\bibinfo
  {journal} {Phys. Rev. A}\ }\textbf {\bibinfo {volume} {80}},\ \bibinfo
  {pages} {023816} (\bibinfo {year} {2009})}\BibitemShut {NoStop}%
\bibitem [{\citenamefont {Tan}\ \emph {et~al.}(2008)\citenamefont {Tan},
  \citenamefont {Erkmen}, \citenamefont {Giovannetti}, \citenamefont {Guha},
  \citenamefont {Lloyd}, \citenamefont {Maccone}, \citenamefont {Pirandola},\
  and\ \citenamefont {Shapiro}}]{Tan2008}%
  \BibitemOpen
  \bibfield  {author} {\bibinfo {author} {\bibfnamefont {S.-H.}\ \bibnamefont
  {Tan}}, \bibinfo {author} {\bibfnamefont {B.~I.}\ \bibnamefont {Erkmen}},
  \bibinfo {author} {\bibfnamefont {V.}~\bibnamefont {Giovannetti}}, \bibinfo
  {author} {\bibfnamefont {S.}~\bibnamefont {Guha}}, \bibinfo {author}
  {\bibfnamefont {S.}~\bibnamefont {Lloyd}}, \bibinfo {author} {\bibfnamefont
  {L.}~\bibnamefont {Maccone}}, \bibinfo {author} {\bibfnamefont
  {S.}~\bibnamefont {Pirandola}}, \ and\ \bibinfo {author} {\bibfnamefont
  {J.~H.}\ \bibnamefont {Shapiro}},\ }\href {\doibase
  10.1103/PhysRevLett.101.253601} {\bibfield  {journal} {\bibinfo  {journal}
  {Phys. Rev. Lett.}\ }\textbf {\bibinfo {volume} {101}},\ \bibinfo {pages}
  {253601} (\bibinfo {year} {2008})}\BibitemShut {NoStop}%
\bibitem [{\citenamefont {Guha}\ and\ \citenamefont {Erkmen}(2009)}]{Guha09}%
  \BibitemOpen
  \bibfield  {author} {\bibinfo {author} {\bibfnamefont {S.}~\bibnamefont
  {Guha}}\ and\ \bibinfo {author} {\bibfnamefont {B.~I.}\ \bibnamefont
  {Erkmen}},\ }\href {\doibase 10.1103/PhysRevA.80.052310} {\bibfield
  {journal} {\bibinfo  {journal} {Phys. Rev. A}\ }\textbf {\bibinfo {volume}
  {80}},\ \bibinfo {pages} {052310} (\bibinfo {year} {2009})}\BibitemShut
  {NoStop}%
\bibitem [{\citenamefont {Brida}\ \emph {et~al.}(2010)\citenamefont {Brida},
  \citenamefont {Genovese},\ and\ \citenamefont {Ruo~Berchera}}]{Brida10}%
  \BibitemOpen
  \bibfield  {author} {\bibinfo {author} {\bibfnamefont {G.}~\bibnamefont
  {Brida}}, \bibinfo {author} {\bibfnamefont {M.}~\bibnamefont {Genovese}}, \
  and\ \bibinfo {author} {\bibfnamefont {I.}~\bibnamefont {Ruo~Berchera}},\
  }\href {\doibase 10.1038/nphoton.2010.29} {\bibfield  {journal} {\bibinfo
  {journal} {Nature Photon.}\ }\textbf {\bibinfo {volume} {4}},\ \bibinfo
  {pages} {227} (\bibinfo {year} {2010})}\BibitemShut {NoStop}%
\bibitem [{\citenamefont {Pirandola}(2011)}]{Pirandola2011a}%
  \BibitemOpen
  \bibfield  {author} {\bibinfo {author} {\bibfnamefont {S.}~\bibnamefont
  {Pirandola}},\ }\href {\doibase 10.1103/PhysRevLett.106.090504} {\bibfield
  {journal} {\bibinfo  {journal} {Phys. Rev. Lett.}\ }\textbf {\bibinfo
  {volume} {106}},\ \bibinfo {pages} {090504} (\bibinfo {year}
  {2011})}\BibitemShut {NoStop}%
\bibitem [{\citenamefont {Sasaki}\ \emph {et~al.}(1997)\citenamefont {Sasaki},
  \citenamefont {Momose},\ and\ \citenamefont {Hirota}}]{sasaki97}%
  \BibitemOpen
  \bibfield  {author} {\bibinfo {author} {\bibfnamefont {M.}~\bibnamefont
  {Sasaki}}, \bibinfo {author} {\bibfnamefont {R.}~\bibnamefont {Momose}}, \
  and\ \bibinfo {author} {\bibfnamefont {O.}~\bibnamefont {Hirota}},\ }\href
  {\doibase 10.1103/PhysRevA.55.3222} {\bibfield  {journal} {\bibinfo
  {journal} {Phys. Rev. A}\ }\textbf {\bibinfo {volume} {55}},\ \bibinfo
  {pages} {3222} (\bibinfo {year} {1997})}\BibitemShut {NoStop}%
\bibitem [{\citenamefont {Paris}(2001)}]{paris01}%
  \BibitemOpen
  \bibfield  {author} {\bibinfo {author} {\bibfnamefont {M.~G.~A.}\
  \bibnamefont {Paris}},\ }\href {\doibase 10.1103/PhysRevA.64.014304}
  {\bibfield  {journal} {\bibinfo  {journal} {Phys. Rev. A}\ }\textbf {\bibinfo
  {volume} {64}},\ \bibinfo {pages} {014304} (\bibinfo {year}
  {2001})}\BibitemShut {NoStop}%
\bibitem [{\citenamefont {Invernizzi}\ \emph {et~al.}(2011)\citenamefont
  {Invernizzi}, \citenamefont {Paris},\ and\ \citenamefont
  {Pirandola}}]{Invernizzi2011}%
  \BibitemOpen
  \bibfield  {author} {\bibinfo {author} {\bibfnamefont {C.}~\bibnamefont
  {Invernizzi}}, \bibinfo {author} {\bibfnamefont {M.~G.~A.}\ \bibnamefont
  {Paris}}, \ and\ \bibinfo {author} {\bibfnamefont {S.}~\bibnamefont
  {Pirandola}},\ }\href {\doibase 10.1103/PhysRevA.84.022334} {\bibfield
  {journal} {\bibinfo  {journal} {Phys. Rev. A}\ }\textbf {\bibinfo {volume}
  {84}},\ \bibinfo {pages} {22334} (\bibinfo {year} {2011})}\BibitemShut
  {NoStop}%
\bibitem [{\citenamefont {Helstrom}(1976)}]{HelstromBook}%
  \BibitemOpen
  \bibfield  {author} {\bibinfo {author} {\bibfnamefont {C.~W.}\ \bibnamefont
  {Helstrom}},\ }\href@noop {} {\emph {\bibinfo {title} {Quantum detection and
  estimation theory}}}\ (\bibinfo  {publisher} {Academic Press New York},\
  \bibinfo {year} {1976})\BibitemShut {NoStop}%
\bibitem [{\citenamefont {Braunstein}\ and\ \citenamefont
  {Caves}(1994)}]{Braunstein94}%
  \BibitemOpen
  \bibfield  {author} {\bibinfo {author} {\bibfnamefont {S.~L.}\ \bibnamefont
  {Braunstein}}\ and\ \bibinfo {author} {\bibfnamefont {C.~M.}\ \bibnamefont
  {Caves}},\ }\href {\doibase 10.1103/PhysRevLett.72.3439} {\bibfield
  {journal} {\bibinfo  {journal} {Phys. Rev. Lett.}\ }\textbf {\bibinfo
  {volume} {72}},\ \bibinfo {pages} {3439} (\bibinfo {year}
  {1994})}\BibitemShut {NoStop}%
\bibitem [{\citenamefont {Paris}(2009)}]{Paris2009}%
  \BibitemOpen
  \bibfield  {author} {\bibinfo {author} {\bibfnamefont {M.~G.~A.}\
  \bibnamefont {Paris}},\ }\href {\doibase 10.1142/S0219749909004839}
  {\bibfield  {journal} {\bibinfo  {journal} {Int. J. Quant. Inf.}\ }\textbf
  {\bibinfo {volume} {7}},\ \bibinfo {pages} {125} (\bibinfo {year}
  {2009})}\BibitemShut {NoStop}%
\bibitem [{\citenamefont {Escher}\ \emph {et~al.}(2011)\citenamefont {Escher},
  \citenamefont {de~Matos~Filho},\ and\ \citenamefont
  {Davidovich}}]{Escher2011}%
  \BibitemOpen
  \bibfield  {author} {\bibinfo {author} {\bibfnamefont {B.~M.}\ \bibnamefont
  {Escher}}, \bibinfo {author} {\bibfnamefont {R.~L.}\ \bibnamefont
  {de~Matos~Filho}}, \ and\ \bibinfo {author} {\bibfnamefont {L.}~\bibnamefont
  {Davidovich}},\ }\href {\doibase 10.1038/nphys1958} {\bibfield  {journal}
  {\bibinfo  {journal} {Nature Phys.}\ }\textbf {\bibinfo {volume} {7}},\
  \bibinfo {pages} {406} (\bibinfo {year} {2011})}\BibitemShut {NoStop}%
\bibitem [{\citenamefont {Monras}\ and\ \citenamefont
  {Paris}(2007)}]{Monras2007}%
  \BibitemOpen
  \bibfield  {author} {\bibinfo {author} {\bibfnamefont {A.}~\bibnamefont
  {Monras}}\ and\ \bibinfo {author} {\bibfnamefont {M.~G.~A.}\ \bibnamefont
  {Paris}},\ }\href {\doibase 10.1103/PhysRevLett.98.160401} {\bibfield
  {journal} {\bibinfo  {journal} {Phys. Rev. Lett.}\ }\textbf {\bibinfo
  {volume} {98}},\ \bibinfo {pages} {160401} (\bibinfo {year}
  {2007})}\BibitemShut {NoStop}%
\bibitem [{\citenamefont {Adesso}\ \emph {et~al.}(2009)\citenamefont {Adesso},
  \citenamefont {Dell'Anno}, \citenamefont {{De Siena}}, \citenamefont
  {Illuminati},\ and\ \citenamefont {Souza}}]{Adesso2009}%
  \BibitemOpen
  \bibfield  {author} {\bibinfo {author} {\bibfnamefont {G.}~\bibnamefont
  {Adesso}}, \bibinfo {author} {\bibfnamefont {F.}~\bibnamefont {Dell'Anno}},
  \bibinfo {author} {\bibfnamefont {S.}~\bibnamefont {{De Siena}}}, \bibinfo
  {author} {\bibfnamefont {F.}~\bibnamefont {Illuminati}}, \ and\ \bibinfo
  {author} {\bibfnamefont {L.~A.~M.}\ \bibnamefont {Souza}},\ }\href {\doibase
  10.1103/PhysRevA.79.040305} {\bibfield  {journal} {\bibinfo  {journal} {Phys.
  Rev. A}\ }\textbf {\bibinfo {volume} {79}},\ \bibinfo {pages} {040305(R)}
  (\bibinfo {year} {2009})}\BibitemShut {NoStop}%
\bibitem [{\citenamefont {{G. Spedalieri, S. L. Braunstein, S.
  Pirandola}}(2016)}]{Gea16}%
  \BibitemOpen
  \bibfield  {author} {\bibinfo {author} {\bibnamefont {{G. Spedalieri, S. L.
  Braunstein, S. Pirandola}}},\ }\href@noop {} {\enquote {\bibinfo {title}
  {{Thermal quantum metrology}},}\ } (\bibinfo {year} {2016}),\ \Eprint
  {http://arxiv.org/abs/1602.05958} {arXiv:1602.05958} \BibitemShut {NoStop}%
\bibitem [{\citenamefont {Monras}\ and\ \citenamefont
  {Illuminati}(2011)}]{Monras2011a}%
  \BibitemOpen
  \bibfield  {author} {\bibinfo {author} {\bibfnamefont {A.}~\bibnamefont
  {Monras}}\ and\ \bibinfo {author} {\bibfnamefont {F.}~\bibnamefont
  {Illuminati}},\ }\href {\doibase 10.1103/PhysRevA.83.012315} {\bibfield
  {journal} {\bibinfo  {journal} {Phys. Rev. A}\ }\textbf {\bibinfo {volume}
  {83}},\ \bibinfo {pages} {012315} (\bibinfo {year} {2011})}\BibitemShut
  {NoStop}%
\bibitem [{\citenamefont {Venzl}\ and\ \citenamefont
  {Freyberger}(2007)}]{Venzl2007}%
  \BibitemOpen
  \bibfield  {author} {\bibinfo {author} {\bibfnamefont {H.}~\bibnamefont
  {Venzl}}\ and\ \bibinfo {author} {\bibfnamefont {M.}~\bibnamefont
  {Freyberger}},\ }\href {\doibase 10.1103/PhysRevA.75.042322} {\bibfield
  {journal} {\bibinfo  {journal} {Phys. Rev. A}\ }\textbf {\bibinfo {volume}
  {75}},\ \bibinfo {pages} {042322} (\bibinfo {year} {2007})}\BibitemShut
  {NoStop}%
\bibitem [{\citenamefont {Crowley}\ \emph {et~al.}(2014)\citenamefont
  {Crowley}, \citenamefont {Datta}, \citenamefont {Barbieri},\ and\
  \citenamefont {Walmsley}}]{Crowley2014}%
  \BibitemOpen
  \bibfield  {author} {\bibinfo {author} {\bibfnamefont {P.~J.~D.}\
  \bibnamefont {Crowley}}, \bibinfo {author} {\bibfnamefont {A.}~\bibnamefont
  {Datta}}, \bibinfo {author} {\bibfnamefont {M.}~\bibnamefont {Barbieri}}, \
  and\ \bibinfo {author} {\bibfnamefont {I.~A.}\ \bibnamefont {Walmsley}},\
  }\href {\doibase 10.1103/PhysRevA.89.023845} {\bibfield  {journal} {\bibinfo
  {journal} {Phys. Rev. A}\ }\textbf {\bibinfo {volume} {89}},\ \bibinfo
  {pages} {023845} (\bibinfo {year} {2014})},\ \Eprint
  {http://arxiv.org/abs/1206.0043} {1206.0043} \BibitemShut {NoStop}%
\bibitem [{\citenamefont {Barbieri}\ \emph {et~al.}(2015)\citenamefont
  {Barbieri}, \citenamefont {Datta}, \citenamefont {Bartley}, \citenamefont
  {Jin}, \citenamefont {Kolthammer},\ and\ \citenamefont
  {Walmsley}}]{Barbieri2015}%
  \BibitemOpen
  \bibfield  {author} {\bibinfo {author} {\bibfnamefont {M.}~\bibnamefont
  {Barbieri}}, \bibinfo {author} {\bibfnamefont {A.}~\bibnamefont {Datta}},
  \bibinfo {author} {\bibfnamefont {T.~J.}\ \bibnamefont {Bartley}}, \bibinfo
  {author} {\bibfnamefont {X.-M.}\ \bibnamefont {Jin}}, \bibinfo {author}
  {\bibfnamefont {W.~S.}\ \bibnamefont {Kolthammer}}, \ and\ \bibinfo {author}
  {\bibfnamefont {I.~A.}\ \bibnamefont {Walmsley}},\ }\href
  {http://arxiv.org/abs/1502.00681} {\enquote {\bibinfo {title} {{Quantum
  enhanced estimation of optical detector efficiencies}},}\ } (\bibinfo {year}
  {2015}),\ \Eprint {http://arxiv.org/abs/1502.00681} {arXiv:1502.00681}
  \BibitemShut {NoStop}%
\bibitem [{\citenamefont {{S. Grandi, A. Zavatta, M. Bellini, M. G. A.
  Paris}}(2015)}]{Grandi2015}%
  \BibitemOpen
  \bibfield  {author} {\bibinfo {author} {\bibnamefont {{S. Grandi, A. Zavatta,
  M. Bellini, M. G. A. Paris}}},\ }\href@noop {} {\enquote {\bibinfo {title}
  {{Experimental quantum tomography of a homodyne detector}},}\ } (\bibinfo
  {year} {2015}),\ \Eprint {http://arxiv.org/abs/1505.03297} {arXiv:1505.03297}
  \BibitemShut {NoStop}%
\bibitem [{\citenamefont {Boyd}(2008)}]{boyd2008nonlinear}%
  \BibitemOpen
  \bibfield  {author} {\bibinfo {author} {\bibfnamefont {R.}~\bibnamefont
  {Boyd}},\ }\href@noop {} {\emph {\bibinfo {title} {{Nonlinear Optics}}}},\
  \bibinfo {edition} {3rd}\ ed.\ (\bibinfo  {publisher} {Academic Press},\
  \bibinfo {address} {Burlington, MA},\ \bibinfo {year} {2008})\BibitemShut
  {NoStop}%
\bibitem [{\citenamefont {Milburn}\ and\ \citenamefont
  {Holmes}(1986)}]{Milburn1986a}%
  \BibitemOpen
  \bibfield  {author} {\bibinfo {author} {\bibfnamefont {G.~J.}\ \bibnamefont
  {Milburn}}\ and\ \bibinfo {author} {\bibfnamefont {C.~A.}\ \bibnamefont
  {Holmes}},\ }\href {\doibase 10.1103/PhysRevLett.56.2237} {\bibfield
  {journal} {\bibinfo  {journal} {Phys. Rev. Lett.}\ }\textbf {\bibinfo
  {volume} {56}},\ \bibinfo {pages} {2237} (\bibinfo {year}
  {1986})}\BibitemShut {NoStop}%
\bibitem [{\citenamefont {Stobińska}\ \emph {et~al.}(2008)\citenamefont
  {Stobińska}, \citenamefont {Milburn},\ and\ \citenamefont
  {W{\'{o}}dkiewicz}}]{Stobinska2008}%
  \BibitemOpen
  \bibfield  {author} {\bibinfo {author} {\bibfnamefont {M.}~\bibnamefont
  {Stobińska}}, \bibinfo {author} {\bibfnamefont {G.~J.}\ \bibnamefont
  {Milburn}}, \ and\ \bibinfo {author} {\bibfnamefont {K.}~\bibnamefont
  {W{\'{o}}dkiewicz}},\ }\href {\doibase 10.1103/PhysRevA.78.013810} {\bibfield
   {journal} {\bibinfo  {journal} {Phys. Rev. A}\ }\textbf {\bibinfo {volume}
  {78}},\ \bibinfo {pages} {013810} (\bibinfo {year} {2008})}\BibitemShut
  {NoStop}%
\bibitem [{\citenamefont {Yurke}\ and\ \citenamefont
  {Stoler}(1986)}]{Yurke1986}%
  \BibitemOpen
  \bibfield  {author} {\bibinfo {author} {\bibfnamefont {B.}~\bibnamefont
  {Yurke}}\ and\ \bibinfo {author} {\bibfnamefont {D.}~\bibnamefont {Stoler}},\
  }\href {\doibase 10.1103/PhysRevLett.57.13} {\bibfield  {journal} {\bibinfo
  {journal} {Phys. Rev. Lett.}\ }\textbf {\bibinfo {volume} {57}},\ \bibinfo
  {pages} {13} (\bibinfo {year} {1986})}\BibitemShut {NoStop}%
\bibitem [{\citenamefont {Yurke}\ and\ \citenamefont
  {Stoler}(1988)}]{Yurke1988}%
  \BibitemOpen
  \bibfield  {author} {\bibinfo {author} {\bibfnamefont {B.}~\bibnamefont
  {Yurke}}\ and\ \bibinfo {author} {\bibfnamefont {D.}~\bibnamefont {Stoler}},\
  }\href {\doibase 10.1016/0378-4363(88)90181-7} {\bibfield  {journal}
  {\bibinfo  {journal} {Physica B}\ }\textbf {\bibinfo {volume} {151}},\
  \bibinfo {pages} {298} (\bibinfo {year} {1988})}\BibitemShut {NoStop}%
\bibitem [{\citenamefont {Miranowicz}\ \emph {et~al.}(1990)\citenamefont
  {Miranowicz}, \citenamefont {Tanas},\ and\ \citenamefont
  {Kielich}}]{Miranowicz2011}%
  \BibitemOpen
  \bibfield  {author} {\bibinfo {author} {\bibfnamefont {A.}~\bibnamefont
  {Miranowicz}}, \bibinfo {author} {\bibfnamefont {R.}~\bibnamefont {Tanas}}, \
  and\ \bibinfo {author} {\bibfnamefont {S.}~\bibnamefont {Kielich}},\ }\href
  {\doibase 10.1088/0954-8998/2/3/006} {\bibfield  {journal} {\bibinfo
  {journal} {Quantum Opt.}\ }\textbf {\bibinfo {volume} {2}},\ \bibinfo {pages}
  {253} (\bibinfo {year} {1990})}\BibitemShut {NoStop}%
\bibitem [{\citenamefont {Paris}(1999)}]{Paris1999}%
  \BibitemOpen
  \bibfield  {author} {\bibinfo {author} {\bibfnamefont {M.~G.~A.}\
  \bibnamefont {Paris}},\ }\href {\doibase 10.1088/1464-4266/1/6/308}
  {\bibfield  {journal} {\bibinfo  {journal} {J. Opt. B}\ }\textbf {\bibinfo
  {volume} {1}},\ \bibinfo {pages} {662} (\bibinfo {year} {1999})}\BibitemShut
  {NoStop}%
\bibitem [{\citenamefont {Jeong}\ \emph {et~al.}(2004)\citenamefont {Jeong},
  \citenamefont {Kim}, \citenamefont {Ralph},\ and\ \citenamefont
  {Ham}}]{Jeong2004}%
  \BibitemOpen
  \bibfield  {author} {\bibinfo {author} {\bibfnamefont {H.}~\bibnamefont
  {Jeong}}, \bibinfo {author} {\bibfnamefont {M.~S.}\ \bibnamefont {Kim}},
  \bibinfo {author} {\bibfnamefont {T.~C.}\ \bibnamefont {Ralph}}, \ and\
  \bibinfo {author} {\bibfnamefont {B.~S.}\ \bibnamefont {Ham}},\ }\href
  {\doibase 10.1103/PhysRevA.70.061801} {\bibfield  {journal} {\bibinfo
  {journal} {Phys. Rev. A}\ }\textbf {\bibinfo {volume} {70}},\ \bibinfo
  {pages} {061801} (\bibinfo {year} {2004})}\BibitemShut {NoStop}%
\bibitem [{\citenamefont {Essiambre}\ and\ \citenamefont
  {Tkach}(2012)}]{Essiambre2012}%
  \BibitemOpen
  \bibfield  {author} {\bibinfo {author} {\bibfnamefont {R.-J.}\ \bibnamefont
  {Essiambre}}\ and\ \bibinfo {author} {\bibfnamefont {R.~W.}\ \bibnamefont
  {Tkach}},\ }\href {\doibase 10.1109/JPROC.2012.2182970} {\bibfield  {journal}
  {\bibinfo  {journal} {Proc. IEEE}\ }\textbf {\bibinfo {volume} {100}},\
  \bibinfo {pages} {1035} (\bibinfo {year} {2012})}\BibitemShut {NoStop}%
\bibitem [{\citenamefont {Luis}(2004)}]{Luis2004}%
  \BibitemOpen
  \bibfield  {author} {\bibinfo {author} {\bibfnamefont {A.}~\bibnamefont
  {Luis}},\ }\href {\doibase 10.1016/j.physleta.2004.06.080} {\bibfield
  {journal} {\bibinfo  {journal} {Phys. Lett. A}\ }\textbf {\bibinfo {volume}
  {329}},\ \bibinfo {pages} {8} (\bibinfo {year} {2004})}\BibitemShut {NoStop}%
\bibitem [{\citenamefont {Rivas}\ and\ \citenamefont {Luis}(2010)}]{Rivas2010}%
  \BibitemOpen
  \bibfield  {author} {\bibinfo {author} {\bibfnamefont {{\'{A}}.}~\bibnamefont
  {Rivas}}\ and\ \bibinfo {author} {\bibfnamefont {A.}~\bibnamefont {Luis}},\
  }\href {\doibase 10.1103/PhysRevLett.105.010403} {\bibfield  {journal}
  {\bibinfo  {journal} {Phys. Rev. Lett.}\ }\textbf {\bibinfo {volume} {105}},\
  \bibinfo {pages} {010403} (\bibinfo {year} {2010})}\BibitemShut {NoStop}%
\bibitem [{\citenamefont {Luis}(2010)}]{Luis2010c}%
  \BibitemOpen
  \bibfield  {author} {\bibinfo {author} {\bibfnamefont {A.}~\bibnamefont
  {Luis}},\ }\href {\doibase 10.1117/6.0000007} {\bibfield  {journal} {\bibinfo
   {journal} {SPIE Rev.}\ }\textbf {\bibinfo {volume} {1}},\ \bibinfo {pages}
  {018006} (\bibinfo {year} {2010})}\BibitemShut {NoStop}%
\bibitem [{\citenamefont {Genoni}\ \emph {et~al.}(2009)\citenamefont {Genoni},
  \citenamefont {Invernizzi},\ and\ \citenamefont {Paris}}]{Genoni2009}%
  \BibitemOpen
  \bibfield  {author} {\bibinfo {author} {\bibfnamefont {M.~G.}\ \bibnamefont
  {Genoni}}, \bibinfo {author} {\bibfnamefont {C.}~\bibnamefont {Invernizzi}},
  \ and\ \bibinfo {author} {\bibfnamefont {M.~G.~A.}\ \bibnamefont {Paris}},\
  }\href {\doibase 10.1103/PhysRevA.80.033842} {\bibfield  {journal} {\bibinfo
  {journal} {Phys. Rev. A}\ }\textbf {\bibinfo {volume} {80}},\ \bibinfo
  {pages} {033842} (\bibinfo {year} {2009})}\BibitemShut {NoStop}%
\bibitem [{\citenamefont {Luis}\ and\ \citenamefont {Rivas}(2015)}]{Luis2015}%
  \BibitemOpen
  \bibfield  {author} {\bibinfo {author} {\bibfnamefont {A.}~\bibnamefont
  {Luis}}\ and\ \bibinfo {author} {\bibfnamefont {{\'{A}}.}~\bibnamefont
  {Rivas}},\ }\href {\doibase 10.1103/PhysRevA.92.022104} {\bibfield  {journal}
  {\bibinfo  {journal} {Phys. Rev. A}\ }\textbf {\bibinfo {volume} {92}},\
  \bibinfo {pages} {022104} (\bibinfo {year} {2015})}\BibitemShut {NoStop}%
\bibitem [{\citenamefont {Cram{\`{e}}r}(1946)}]{Cramer1946}%
  \BibitemOpen
  \bibfield  {author} {\bibinfo {author} {\bibfnamefont {H.}~\bibnamefont
  {Cram{\`{e}}r}},\ }\href@noop {} {\emph {\bibinfo {title} {{Mathematical
  Methods of Statistics}}}}\ (\bibinfo  {publisher} {Princeton Univ. Press},\
  \bibinfo {address} {Princeton},\ \bibinfo {year} {1946})\BibitemShut
  {NoStop}%
\bibitem [{\citenamefont {D'Ariano}(1994)}]{DAriano1994}%
  \BibitemOpen
  \bibfield  {author} {\bibinfo {author} {\bibfnamefont {G.~M.}\ \bibnamefont
  {D'Ariano}},\ }\href {\doibase 10.1016/0375-9601(94)90899-0} {\bibfield
  {journal} {\bibinfo  {journal} {Phys. Lett. A}\ }\textbf {\bibinfo {volume}
  {187}},\ \bibinfo {pages} {231} (\bibinfo {year} {1994})}\BibitemShut
  {NoStop}%
\bibitem [{\citenamefont {Milburn}\ \emph {et~al.}(1989)\citenamefont
  {Milburn}, \citenamefont {Mecozzi},\ and\ \citenamefont
  {Tombesi}}]{Milburn1989}%
  \BibitemOpen
  \bibfield  {author} {\bibinfo {author} {\bibfnamefont {G.}~\bibnamefont
  {Milburn}}, \bibinfo {author} {\bibfnamefont {A.}~\bibnamefont {Mecozzi}}, \
  and\ \bibinfo {author} {\bibfnamefont {P.}~\bibnamefont {Tombesi}},\ }\href
  {\doibase 10.1080/09500348914551721} {\bibfield  {journal} {\bibinfo
  {journal} {J. Mod. Opt.}\ }\textbf {\bibinfo {volume} {36}},\ \bibinfo
  {pages} {1607} (\bibinfo {year} {1989})}\BibitemShut {NoStop}%
\bibitem [{\citenamefont {Pe\v{r}inov{\'{a}}}\ and\ \citenamefont
  {Luk{\v{s}}}(1988)}]{Perinova1988}%
  \BibitemOpen
  \bibfield  {author} {\bibinfo {author} {\bibfnamefont {V.}~\bibnamefont
  {Pe\v{r}inov{\'{a}}}}\ and\ \bibinfo {author} {\bibfnamefont
  {A.}~\bibnamefont {Luk{\v{s}}}},\ }\href {\doibase 10.1080/09500348814551631}
  {\bibfield  {journal} {\bibinfo  {journal} {J. Mod. Opt.}\ }\textbf {\bibinfo
  {volume} {35}},\ \bibinfo {pages} {1513} (\bibinfo {year}
  {1988})}\BibitemShut {NoStop}%
\bibitem [{\citenamefont {Pe\v{r}inov{\'{a}}}\ and\ \citenamefont
  {Luk{\v{s}}}(1990)}]{Peinova1990}%
  \BibitemOpen
  \bibfield  {author} {\bibinfo {author} {\bibfnamefont {V.}~\bibnamefont
  {Pe\v{r}inov{\'{a}}}}\ and\ \bibinfo {author} {\bibfnamefont
  {A.}~\bibnamefont {Luk{\v{s}}}},\ }\href {\doibase 10.1103/PhysRevA.41.414}
  {\bibfield  {journal} {\bibinfo  {journal} {Phys. Rev. A}\ }\textbf {\bibinfo
  {volume} {41}},\ \bibinfo {pages} {414} (\bibinfo {year} {1990})}\BibitemShut
  {NoStop}%
\bibitem [{\citenamefont {Chaturvedi}\ and\ \citenamefont
  {Srinivasan}(1991)}]{Chaturvedi1991}%
  \BibitemOpen
  \bibfield  {author} {\bibinfo {author} {\bibfnamefont {S.}~\bibnamefont
  {Chaturvedi}}\ and\ \bibinfo {author} {\bibfnamefont {V.}~\bibnamefont
  {Srinivasan}},\ }\href {\doibase 10.1080/09500349114550761} {\bibfield
  {journal} {\bibinfo  {journal} {J. Mod. Opt.}\ }\textbf {\bibinfo {volume}
  {38}},\ \bibinfo {pages} {777} (\bibinfo {year} {1991})}\BibitemShut
  {NoStop}%
\bibitem [{\citenamefont {Bina}\ \emph {et~al.}(2014)\citenamefont {Bina},
  \citenamefont {Mandarino}, \citenamefont {Olivares},\ and\ \citenamefont
  {Paris}}]{Bina2014}%
  \BibitemOpen
  \bibfield  {author} {\bibinfo {author} {\bibfnamefont {M.}~\bibnamefont
  {Bina}}, \bibinfo {author} {\bibfnamefont {A.}~\bibnamefont {Mandarino}},
  \bibinfo {author} {\bibfnamefont {S.}~\bibnamefont {Olivares}}, \ and\
  \bibinfo {author} {\bibfnamefont {M.~G.~A.}\ \bibnamefont {Paris}},\ }\href
  {\doibase 10.1103/PhysRevA.89.012305} {\bibfield  {journal} {\bibinfo
  {journal} {Phys. Rev. A}\ }\textbf {\bibinfo {volume} {89}},\ \bibinfo
  {pages} {012305} (\bibinfo {year} {2014})}\BibitemShut {NoStop}%
\bibitem [{\citenamefont {Mandarino}\ \emph {et~al.}(2014)\citenamefont
  {Mandarino}, \citenamefont {Bina}, \citenamefont {Olivares},\ and\
  \citenamefont {Paris}}]{Mandarino2014}%
  \BibitemOpen
  \bibfield  {author} {\bibinfo {author} {\bibfnamefont {A.}~\bibnamefont
  {Mandarino}}, \bibinfo {author} {\bibfnamefont {M.}~\bibnamefont {Bina}},
  \bibinfo {author} {\bibfnamefont {S.}~\bibnamefont {Olivares}}, \ and\
  \bibinfo {author} {\bibfnamefont {M.~G.~A.}\ \bibnamefont {Paris}},\ }\href
  {\doibase 10.1142/S0219749914610152} {\bibfield  {journal} {\bibinfo
  {journal} {Int. J. Quant. Inf.}\ }\textbf {\bibinfo {volume} {12}},\ \bibinfo
  {pages} {1461015} (\bibinfo {year} {2014})}\BibitemShut {NoStop}%
\bibitem [{\citenamefont {Genoni}\ and\ \citenamefont
  {Paris}(2010)}]{Genoni2010}%
  \BibitemOpen
  \bibfield  {author} {\bibinfo {author} {\bibfnamefont {M.~G.}\ \bibnamefont
  {Genoni}}\ and\ \bibinfo {author} {\bibfnamefont {M.~G.~A.}\ \bibnamefont
  {Paris}},\ }\href {\doibase 10.1103/PhysRevA.82.052341} {\bibfield  {journal}
  {\bibinfo  {journal} {Phys. Rev. A}\ }\textbf {\bibinfo {volume} {82}},\
  \bibinfo {pages} {052341} (\bibinfo {year} {2010})}\BibitemShut {NoStop}%
\end{thebibliography}%
\end{document}